\documentclass[prd,aps,nofootinbib,floatfix,10pt]{revtex4}

\usepackage{amsmath,graphicx,epsfig,amssymb,dsfont,mathtools,}
\usepackage[usenames]{color}
\usepackage{ulem} 
\usepackage{bigstrut}
\usepackage{multirow}
\usepackage{makecell}
\usepackage{verbatim}
\usepackage{amsmath,graphicx,color,epsfig}
  \allowdisplaybreaks[1]

\allowdisplaybreaks
\renewcommand{\thesubsection}{\Roman{subsection}}

\begin{document}
\title{Weak Decays of  Triply Heavy Tetraquarks ${b\bar c}{b\bar q}$ }
\author{Ye Xing~\footnote{Email:xingye\_guang@cumt.edu.cn}}
\affiliation{
School of Physics,\\ China University of Mining and Technology, Xuzhou 221000, China }

\begin{abstract}
We study lifetimes and weak decays of the triply heavy tetraquarks ${b\bar c}{b\bar q}$. Following the heavy quark expanding(HQE), the lifetimes of tetraquarks ${b\bar c}{b\bar q}$ can be expressed as the summation of different dimension operators. Particularly, we obtain the lifetimes of ${b\bar c}{b\bar q}$ at the next-to-leading order(NLO) given as $\tau(T^{\{bb\}}_{\{\bar c\bar q\}})=0.70\times 10^{-12}s$. Besides, we construct the weak decays Hamiltonian of ${b\bar c}{b\bar q}$ in hadronic level under the SU(3) flavor symmetry. The discussion of the Hamiltonian can deduce the decay amplitudes and width relations of the tetraquarks. Following the choosing rules, we collect some golden channels for the mesonic decays of tetraquarks, which will be helpful to search for triply heavy tetraquarks ${b\bar c}{b\bar q}$ in future experiments.
\end{abstract}

\maketitle

\section{Introduction}

The quark model could well explain the hadronic states for a long time, in which mesons are handled as the bound states of a constituent quark and anti-quark, and baryons are bound by three constituent quarks. Nevertheless, the Belle collaboration changes the situation in 2003 for the  discovery of the X(3872)~\cite{Choi:2003ue}, in $B^{\pm}\to K^{\pm} X(X\to \pi^+\pi^- J/\psi)$ decays. The new state which contains the $c\bar c$ pair is not match with the ordinary quarkonium state, consequently expected to be the candidate of four quark exotic state~\cite{Ader:1981db,Maiani:2004vq,Maiani:2018tfe,Agaev:2016ifn,Agaev:2018vag, Guo:2017jvc,Guo:2013sya,Cleven:2013sq,Guo:2013ufa,Liu:2016xly,Li:2013xia,Guo:2014ppa,Guo:2014iya,Chen:2016mjn,Wang:2013kra,Li:2014uia,Albaladejo:2017blx,Guo:2013zbw,Voloshin:2013ez,Chen:2011pv,Chen:2011pu,Bondar:2011ev,Li:2015uwa,Chen:2013bha,Wang:2018poa,Brambilla:2019esw}. Subsequently, many more exotic states are observed in processes, for instance, BES-III and Belle collaborations observe charged heavy quarkoniumlike states $Z_c(3900)^{\pm}$~\cite{Ablikim:2013mio,Liu:2013dau,Xiao:2013iha,Ablikim:2015tbp}, $Z_c(4020)^{\pm}$~\cite{Ablikim:2013wzq,Ablikim:2014dxl}, $Z_b(10610)^{\pm}$~\cite{Belle:2011aa} and $Z_b(10650)^{\pm}$~\cite{Belle:2011aa}. Until now, the X(3872) is one of the best studied four quark exotic state candidates, whose mass right at the $D^0 \bar D^{*0}$ threshold, and spin-parity quantum numbers $J^{PC}=1^{++}$~\cite{Aaij:2015eva}. However, there is still a open issue whether the state is a loosely bound molecule or a compact tetraquark.

The four-quark state $b\bar cb\bar q$ with three heavy quark is apparently different from the discovered quarkoniumlike state, in this respect, it might offer a new platform to study the internal structure of the exotic states. In addition, it can also be an ideal source to understand the hadronic dynamic and  QCD factorization approach. Some theoretical and experimental studies have been carried out to explore the properties of the exotic state  at present, especially the mass spectrum  by the simple quark model~\cite{Chen:2016ont}, QCD sum rule~\cite{Jiang:2017tdc}. In this paper, we will concern on the lifetime and decays channels of the state. Since the experiment give no conclusion about the state, it is believed that the four-quark state might be the stable tetraquarks whose dominant decay modes would be induced by the weak interaction. In the diquark-antidiquark model~\cite{Jaffe:2003sg}, the lifetimes of lowest lying tetraquarks with the spin-parity $0^+$ and $1^+$ can be achieved by the operator product expansion(OPE) technique~\cite{Lenz:2015dra,Ali:2018xfq}. Further more, the flavor SU(3) symmetry analysis which have been successfully applied into heavy mesons or baryons~\cite{Jaffe:2004ph,Savage:1989ub,Gronau:1995hm,He:1998rq,Chiang:2004nm,Li:2007bh,Wang:2009azc,Cheng:2011qh,Hsiao:2015iiu,Lu:2016ogy,He:2016xvd,Wang:2017vnc,Wang:2017azm,Wang:2017mqp,Shi:2017dto,He:2018php,Wang:2018utj} can be used to study the weak decays of $b\bar cb\bar q$.

Generally, the lifetimes of the tetraquarks can be expressed into several matrix elements of effective operators in the OPE technique. In this case, we investigate the lifetimes at the leading and next-to-leading order in the heavy quark expansion.  In the other case, the weak decays of the tetraquarks can be considered at the hadronic level Hamiltonian in the SU(3) symmetry. The triply heavy tetraquarks $b\bar cb\bar q$ and the heavy quark transition can be classified in the SU(3) light quark symmetry. Thus one can construct the Hamiltonian in the hadron level, and the non-perturbative effect can be absorbed into some parameters $a_i,b_i,c_i,\ldots$. The discussion of the Hamiltonian can tell us the decay amplitudes and relations between different channels. Such analyses are benefit to the searching for the triply heavy tetraquarks in future experiment.

The paper is organized as follows. In Sec.II, we will give the particle multiplets in the SU(3) symmetry. Section III concerns on the lifetimes of the tetraqurak ${b\bar c}{b\bar q}$ with the approach of OPE. The possible weak decays of the tetraqurak ${b\bar c}{b\bar q}$ are discussed in Section IV, including mesonic two-body non-leptonic decays and the three- or four-body semi-leptonic decays. In Section V, we present a collection of golden decay channels. The summary is given in the end.

\section{Particle Multiplets}
\label{sec:particle_multiplet}
\begin{figure}
  \centering
  \includegraphics[width=0.88\columnwidth]{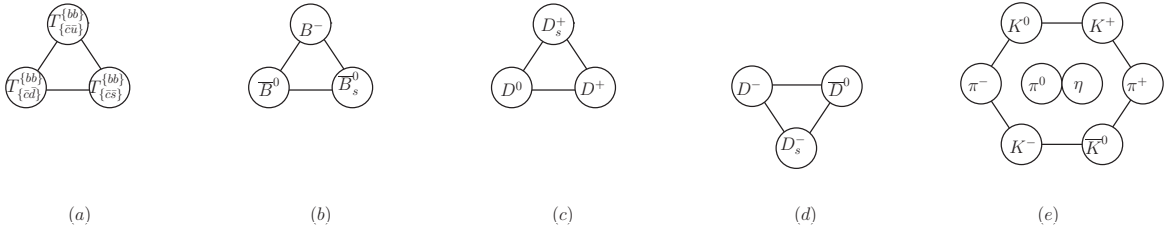}\\
  \caption{The weight diagrams for the tetraquark $T_{\{\bar c \bar q\}}^{\{bb\}}$, bottom meson and charmed meson anti-triplet, anti-charmed meson triplet and light meson octet.}\label{fig:weight1}
\end{figure}

The tetraquark with the quark constituents ${bb}{\bar c\bar q}$ is a flavor SU(3) anti-triplet signed as $T_{\{\bar c \bar q\}}^{\{bb\}}$. In the following, we will concern on the mesonic decays of $T_{\{\bar c \bar q\}}^{\{bb\}}$. Therefore, the particle multiplets are given firstly. Following the SU(3) analysis, we give the SU(3) representation of the tetraquark $T_{\{\bar c \bar q\}}^{\{bb\}}$,

\begin{eqnarray}
 T_{\{\bar c \bar q\}}^{\{bb\}}= \Big(T_{\{\bar c \bar u\}}^{\{bb\}},T_{\{\bar c \bar d\}}^{\{bb\}},T_{\{\bar c \bar s\}}^{\{bb\}}\Big).
\end{eqnarray}
The light pseudoscalar mesons can form an octet, which represented as
\begin{eqnarray}
 M_{8}=\begin{pmatrix}
 \frac{\pi^0}{\sqrt{2}}+\frac{\eta}{\sqrt{6}}
 &\pi^+ & K^+\\
 \pi^-&-\frac{\pi^0}{\sqrt{2}}+\frac{\eta}{\sqrt{6}}&{K^0}\\
 K^-&\bar K^0 &-2\frac{\eta}{\sqrt{6}}
 \end{pmatrix}.
\end{eqnarray}
Besides, we give the SU(3) representations of bottom and charmed mesons,
\begin{eqnarray}
B_i=\left(\begin{array}{ccc} B^-, & \overline B^0, &\overline B^0_s  \end{array} \right),
D_i=\left(\begin{array}{ccc} D^0, & D^+, & D^+_s  \end{array} \right), \;\;\;
\overline D^i=\left(\begin{array}{ccc}\overline D^0, & D^-, & D^-_s  \end{array} \right).
\end{eqnarray}
For completeness, we show the weight diagrams of the states above given in Fig.~\ref{fig:weight1}.

\section{LIFETIMES}
\label{sec:particle_multiplet}
In the section, we will study the lifetimes of the tetraquark $T_{\{\bar c \bar q\}}^{\{bb\}}$ under the OPE. We proceed as follows,
the decays width of $T^{\{bb\}}_{\{\bar c\bar q\}}\to X$ with the spin-parity $J^P=(0^+,1^+)$ can be expressed respectively as
\begin{eqnarray}
&\Gamma(T^{\{bb\}}_{\{\bar c\bar q\}}(0^+)\to X)=\frac{1}{2m_T} \sum_X \int \prod_i \Big[ \frac{d^3 \overrightarrow{p}_i}{(2\pi)^3 2E_i}\Big] (2\pi)^4 \delta^4(p_T-\sum_i p_i) \sum_{\lambda}|\langle X|\mathcal{H}| T^{\{bb\}}_{\{\bar c\bar q\}}\rangle|^2,\\
&\Gamma(T^{\{bb\}}_{\{\bar c\bar q\}}(1^+)\to X)=\frac{1}{2m_T} \sum_X \int \prod_i \Big[ \frac{d^3 \overrightarrow{p}_i}{(2\pi)^3 2E_i}\Big] (2\pi)^4 \delta^4(p_T-\sum_i p_i) \frac{1}{3} \sum_{\lambda}|\langle X|\mathcal{H}| T^{\{bb\}}_{\{\bar c\bar q\}}\rangle|^2,
\end{eqnarray}

here, $m_T$, $\lambda$ and $p_T^{\mu}$ are the mass, spin and four-momentum  of tetraquark $T_{\{\bar c \bar q\}}^{\{bb\}}$ respectively.
The matrix in full theory Hamiltonian $\mathcal{H}$ can match with that of electro-weak effective Hamiltonian $\mathcal{H}_{eff}^{ew}$, whose Hamiltonian given as
\begin{eqnarray}
\mathcal{H}_{eff}^{ew}=\frac{G_F}{\sqrt{2}} \Big[ \sum_{q=u,c} V_c^q(C_1 Q_1^q+C_2 Q_2^q)-V_p \sum_{j=3} C_j O_j\Big],
\end{eqnarray}
where $C_i$ and $O_i$ are Wilson coefficient and effective operator. Besides, $V$s are the combinations of Cabibbo-Kobayashi-Maskawa(CKM) elements.
The total decay width of $\Gamma(T^{\{bb\}}_{\{\bar c\bar q\}}\to X)$ can be deduced in the optical theorem, rewritten as
\begin{eqnarray}
\Gamma(T^{\{bb\}}_{\{\bar c\bar q\}}\to X)=\frac{1}{2m_T}\sum_{\lambda}\langle T^{\{bb\}}_{\{\bar c\bar q\}}|\mathcal{T}|T^{\{bb\}}_{\{\bar c\bar q\}}\rangle,\\
\mathcal{T}=Im\ i \int d^4x T\{ \mathcal{H}_{eff}(x) \mathcal{H}_{eff}(0)\}.
\end{eqnarray}
Further more, the transition operator can be expanded in the heavy quark expanding(HQE), which given as
\begin{eqnarray}
\mathcal{T}=\sum_{Q=b,c}\frac{G_F^2m_Q^5}{192\pi^3} |V_{CKM}|^2 \Big[c_{3,Q}(\bar QQ)+\frac{c_{5,Q}}{m_Q^2}(\bar Q g_s \sigma_{\mu\nu}G^{\mu\nu}Q)+2\frac{c_{6,Q}}{m_Q^3}(\bar Qq)_{\Gamma}(\bar q Q)_{\Gamma} \Big],
\end{eqnarray}
here, $G_F$ is the Fermi constant and $V_{CKM}$ is the CKM element. The coefficients $c_{i,Q}$ which coming from the heavy quark decays are the perturbative coefficients. Therefore, the total decay width of the tetraquark $T_{\{\bar c \bar q\}}^{\{bb\}}$ can be given in the leading dimension contribution as
\begin{eqnarray}
&\Gamma(T^{\{bb\}}_{\{\bar c\bar q\}}(0^+) \to X)= \sum_{Q=b,c} \frac{G_F^2m_Q^5}{192\pi^3} |V_{CKM}|^2 \sum_{\lambda} \frac{\langle T^{\{bb\}}_{\{\bar c\bar q\}}|\bar Q Q|T^{\{bb\}}_{\{\bar c\bar q\}}\rangle}{2m_T},
\end{eqnarray}
where the heavy quark matrix elements are corresponding with bottom and charm number in the tetraquark state.
\begin{eqnarray}
\sum_{\lambda} \frac{\langle T^{\{bb\}}_{\{\bar c\bar q\}}|\bar b b|T^{\{bb\}}_{\{\bar c\bar q\}}\rangle}{2m_T}=2+\mathcal{O}(1/m_b),\; \sum_{\lambda} \frac{\langle T^{\{bb\}}_{\{\bar c\bar q\}}|\bar c c|T^{\{bb\}}_{\{\bar c\bar q\}}\rangle}{2m_T}=1+\mathcal{O}(1/m_c).
\end{eqnarray}
The perturbative short-distance coefficients $c_{3,Q}$s have been determined as $c_{3,b}=5.29\pm0.35,\ c_{3,c}=6.29\pm0.72$ at the leading order(LO) and $c_{3,b}=6.88\pm0.74,\ c_{3,c}=11.61\pm1.55$ at the next-to-leading order(NLO)~\cite{Lenz:2015dra}. Accordingly, we expect the lifetimes of the tetraquark given as
\begin{eqnarray}
&\Gamma(T^{\{bb\}}_{\{\bar c\bar q\}}(0^+))=\left\{\begin{array}{l} (1.85\pm0.16)\times 10^{-12} \ {\rm GeV} ,
\;\text{LO} \\ (2.83\pm0.34)\times 10^{-12} \ {\rm GeV}  ,\; \text{NLO} \end{array}\right. ,
&\Gamma(T^{\{bb\}}_{\{\bar c\bar q\}}(1^+))=\left\{\begin{array}{l} (6.18\pm0.53)\times 10^{-13} \ {\rm GeV} ,
\;\text{LO} \\ (9.43\pm1.13)\times 10^{-13} \ {\rm GeV}  ,\; \text{NLO} \end{array}\right. ,\\
&\tau(T^{\{bb\}}_{\{\bar c\bar q\}}(0^+))=\left\{\begin{array}{l} (3.21\pm0.27)\times 10^{-12} \ s ,
\;\text{LO} \\ (2.10\pm0.27)\times 10^{-12} \ s  ,\; \text{NLO} \end{array}\right. ,
&\tau(T^{\{bb\}}_{\{\bar c\bar q\}}(1^+))=\left\{\begin{array}{l} (1.07\pm0.09)\times 10^{-12} \ s ,
\;\text{LO} \\ (0.70\pm0.09)\times 10^{-12} \ s  ,\; \text{NLO} \end{array}\right. ,
\end{eqnarray}
where the heavy quark masses $m_c=1.4\ {\rm GeV}$ and $m_b=4.8\ {\rm GeV}$.

\section{Weak decays}
\renewcommand\thesubsection{(\Roman{subsection})}
\label{sec:weak_decay}
In this section, we will study the possible weak decay of the tetraquark $T_{\{\bar c\bar q\}}^{\{bb\}}$. Generally, we can classify the decays modes by the quantities of CKM matrix elements.

\begin{itemize}

\item For the $b/c$ quark semi-leptonic decays, we adopt the following groups.
\begin{eqnarray}
b\to c/u \ell^- \bar \nu_{\ell},\ \ \
  \bar c\to  \bar d/\bar s  \ell^-   \bar \nu_{\ell}.
\end{eqnarray}
The general electro-weak  Hamiltonian can be expressed as
\begin{eqnarray}
 {\cal H}_{eff} &=& \frac{G_F}{\sqrt2} \left[V_{q'b} \bar q' \gamma^\mu(1-\gamma_5)b \bar  \ell\gamma_\mu(1-\gamma_5) \nu_{\ell}+V_{cq} \bar c  \gamma^\mu(1-\gamma_5)q \bar \ell \gamma_\mu(1-\gamma_5) \nu_{\ell}\right] +h.c.,
\end{eqnarray}
where $q'=(u,c)$, $q=(d,s)$. The operators of $b\to u/c \ell^- \bar \nu_{\ell}$ transition form an SU(3) flavor triplet $H_{3}'$ or singlet. In addition, the transition $\bar c\to \bar q \ell^-\bar \nu$ can form a triplet $H_{  3}$.

\item  For the $\bar c$ quark non-leptonic decays, the groups are given as
\begin{eqnarray}
\bar c\to \bar s d \bar u, \;
\bar c\to \bar u d \bar d/s \bar s, \;
\bar c\to \bar d s \bar u, \;
\end{eqnarray}
which are Cabibbo allowed, singly Cabibbo suppressed and doubly Cabibbo suppressed respectively.
The operator $\bar c q_1  \bar q_2 q_3$ of the transition transforms
 as ${\bf  \bar3}\otimes {\bf 3}\otimes {\bf
\bar3}={\bf  \bar3}\oplus {\bf  \bar3}\oplus {\bf6}\oplus {\bf \overline {15}}$, marked as $H_{ 3},H_{ 6}$ and $H_{ \overline {15}}$.

\item For the b quark non-leptonic decays, the groups are classified as
\begin{eqnarray}
b\to c\bar c d/s, \;
b\to c \bar u d/s, \;
b\to u \bar c d/s, \;
b\to q_1 \bar q_2 q_3,
\end{eqnarray}
$q_{1,2,3}$ represent the light quark($d/s$). The operator of the transition $b\to c\bar c d/s$ form  an triplet $H_{  3}$, and the operator of the transition $b\to c \bar u d/s$  can form an octet $H_{{8}}$. In addition, the operator of transition $b\to u\bar cs/d$ can form an anti-symmetric ${\bf  \bar 3}$ plus a symmetric ${\bf  6}$ tensors.
While the charmless tree level operator $(\bar q_1 b)(\bar q_2 q_3)$ ($q_i=d,s$) can be decomposed as $\bf3\otimes\bar {\bf3}\otimes\bf3=\bf3\oplus\bf3\oplus \bf{\bar 6}\oplus \bf{15}$, they are signed as $H_{ 3},H_{ \overline 6}$ and $H_{ 15}$.
\end{itemize}

The parameters of the tensors above can be found in Ref.~\cite{Shi:2017dto,Li:2019uch,Li:2018bkh,Xing:2018bqt}, such as $(H_{ 6})_{31}^2=-(H_{6})_{13}^2=1$ for the transition of $\bar c\to \bar s  d \bar u$. In the following, we will study the possible weak decay modes of  $T^{\{bb\}}_{\{\bar c\bar q\}}$ in order.

\subsection{Semi-Leptonic decays}

\begin{figure}
  \centering
  \includegraphics[width=0.99\columnwidth]{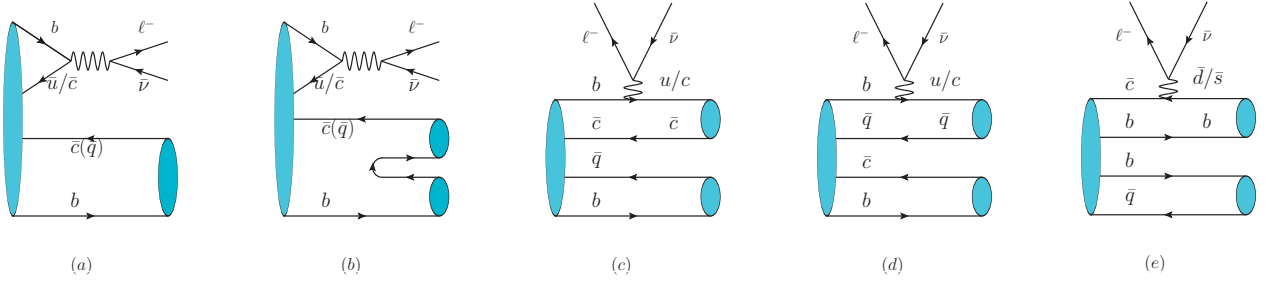}\\
  \caption{The Feynman diagrams for the tetraquark $T^{\{bb\}}_{\{\bar c\bar q\}}$ semi-leptonic decays into mesons. The panel(a) represents the one meson in final states.  The panel(b-e) show the processes of two mesons in final states.}\label{fig:topology1}
\end{figure}

Following the SU(3) analysis, we respectively construct the Hamiltonian of hadronic level for the semi-leptonic decays  $b\to c/u \ell^- \overline \nu_{\ell}$,
\begin{eqnarray}
&\mathcal{H}_{eff}&=a_1 (T^{\{bb\}}_{\{\bar c\bar q\}})_i (H_3')^i \overline{B}_c \bar \ell \nu+a_2 (T^{\{bb\}}_{\{\bar c\bar q\}})_i (H_3')^i D_j \overline{B}^j \bar \ell \nu +a_3 (T^{\{bb\}}_{\{\bar c\bar q\}})_i (H_3')^j D_j \overline{B}^i \bar \ell \nu\nonumber\\
&&+a_4 (T^{\{bb\}}_{\{\bar c\bar q\}})_i (H_3')^j M_j^i \overline{B}_c \bar \ell \nu,\label{eq:semil1}\\
&\mathcal{H}_{eff}&=a_5 (T^{\{bb\}}_{\{\bar c\bar q\}})_i \overline{B}^i \bar \ell \nu+a_6 (T^{\{bb\}}_{\{\bar c\bar q\}})_i  \overline{B}^j M^i_j\bar \ell \nu+a_7 (T^{\{bb\}}_{\{\bar c\bar q\}})_i \overline{B}^i J/\psi \bar \ell \nu +a_8 (T^{\{bb\}}_{\{\bar c\bar q\}})_i  \overline{D}^i \overline{B_c}\bar \ell \nu.\label{eq:semil2}
\end{eqnarray}
The corresponding Feynman diagrams are shown in Fig.~\ref{fig:topology1}.(a-d). Further more, the diagram Fig.~\ref{fig:topology1}.(a) is relative to the $a_1$ and $a_5$ terms. The diagrams Fig.~\ref{fig:topology1}.(b-d) of four-body semi-leptonic decays  are  respectively corresponding with the remanding terms in the Hamiltonian Eq.~\eqref{eq:semil1} and Eq.~\eqref{eq:semil2}.
Expanding the Hamiltonian, we can obtain the decay amplitudes of different decay channels, which are collected in Tab.~\ref{tab:bq_Bc_lv}. The relations of decay widths can be further deduced, when the phase space effect is neglected.
\begin{eqnarray*}
&&\Gamma(T^{\{bb\}}_{\{\bar c\bar u\}}\to  \overline D^0 B^-l^-\bar\nu)=
    { }\Gamma(T^{\{bb\}}_{\{\bar c\bar u\}}\to  D^- \overline B^0l^-\bar\nu)=
    \Gamma(T^{\{bb\}}_{\{\bar c\bar u\}}\to   D^-_s \overline B^0_sl^-\bar\nu),\\&&
    \Gamma(T^{\{bb\}}_{\{\bar c\bar u\}}\to \pi^0   B_c^- l^-\bar\nu)=
3\Gamma(T^{\{bb\}}_{\{\bar c\bar u\}}\to \eta   B_c^- l^-\bar\nu)=
\frac{1}{2}\Gamma(T^{\{bb\}}_{\{\bar c\bar d\}}\to \pi^+   B_c^- l^-\bar\nu)=
\frac{1}{2}\Gamma(T^{\{bb\}}_{\{\bar c\bar s\}}\to K^+   B_c^- l^-\bar\nu).\\
&&\Gamma(T^{\{bb\}}_{\{\bar c\bar u\}}\to B^- l^-\bar\nu)=
\Gamma(T^{\{bb\}}_{\{\bar c\bar d\}}\to \overline B^0 l^-\bar\nu)=
\Gamma(T^{\{bb\}}_{\{\bar c\bar s\}}\to \overline B^0_s l^-\bar\nu),\\
&&\Gamma(T^{\{bb\}}_{\{\bar c\bar u\}}\to B^-  J/\psi l^-\bar\nu)=
\Gamma(T^{\{bb\}}_{\{\bar c\bar d\}}\to \overline B^0  J/\psi l^-\bar\nu)=
\Gamma(T^{\{bb\}}_{\{\bar c\bar s\}}\to \overline B^0_s  J/\psi l^-\bar\nu),\\
&&\Gamma(T^{\{bb\}}_{\{\bar c\bar u\}}\to  D^0  B_c^- l^-\bar\nu)=
\Gamma(T^{\{bb\}}_{\{\bar c\bar d\}}\to  D^+  B_c^- l^-\bar\nu)=
\Gamma(T^{\{bb\}}_{\{\bar c\bar s\}}\to  D^+_s  B_c^- l^-\bar\nu),\\
&&\Gamma(T^{\{bb\}}_{\{\bar c\bar u\}}\to  B^- \pi^0 l^-\bar\nu)= 3\Gamma(T^{\{bb\}}_{\{\bar c\bar u\}}\to  B^- \eta l^-\bar\nu)=\frac{1}{2}\Gamma(T^{\{bb\}}_{\{\bar c\bar u\}}\to  \overline B^0 \pi^- l^-\bar\nu)=\frac{1}{2}\Gamma(T^{\{bb\}}_{\{\bar c\bar u\}}\to  \overline B^0_s K^- l^-\bar\nu)\\
&&=\frac{1}{2}\Gamma(T^{\{bb\}}_{\{\bar c\bar d\}}\to  B^- \pi^+ l^-\bar\nu)=\Gamma(T^{\{bb\}}_{\{\bar c\bar d\}}\to  \overline B^0 \pi^0 l^-\bar\nu)=3\Gamma(T^{\{bb\}}_{\{\bar c\bar d\}}\to  \overline B^0 \eta l^-\bar\nu)=\frac{1}{2}\Gamma(T^{\{bb\}}_{\{\bar c\bar d\}}\to  \overline B^0_s \overline K^0 l^-\bar\nu)\\
&&=\frac{1}{2}\Gamma(T^{\{bb\}}_{\{\bar c\bar s\}}\to  B^- K^+ l^-\bar\nu)=\frac{1}{2}\Gamma(T^{\{bb\}}_{\{\bar c\bar s\}}\to  \overline B^0 K^0 l^-\bar\nu)=\frac{3}{4}\Gamma(T^{\{bb\}}_{\{\bar c\bar s\}}\to  \overline B^0_s \eta l^-\bar\nu).
\end{eqnarray*}

\begin{table}
\caption{Amplitudes for tetraquark $T^{\{bb\}}_{\{\bar{c} \bar{q}\}}$ semi-leptonic decays into mesons with the transition of $b\to c/u \ell^- \overline \nu_{\ell}$.}\label{tab:bq_Bc_lv}\begin{tabular}{|c|c|c|c|c|c|c|c}\hline\hline
channel & amplitude &channel &amplitude\\\hline
$T^{\{bb\}}_{\{\bar c\bar u\}}\to B_c^- l^-\bar\nu $ & $ a_1 V_{\text{ub}}$&&\\\hline
\hline
$T^{\{bb\}}_{\{\bar c\bar u\}}\to \pi^0   B_c^- l^-\bar\nu $ & $ \frac{a_4 V_{\text{ub}}}{\sqrt{2}}$&
$T^{\{bb\}}_{\{\bar c\bar u\}}\to \eta   B_c^- l^-\bar\nu $ & $ \frac{a_4 V_{\text{ub}}}{\sqrt{6}}$\\\hline
$T^{\{bb\}}_{\{\bar c\bar d\}}\to \pi^+   B_c^- l^-\bar\nu $ & $ a_4 V_{\text{ub}}$&
$T^{\{bb\}}_{\{\bar c\bar s\}}\to K^+   B_c^- l^-\bar\nu $ & $ a_4 V_{\text{ub}}$\\\hline
\hline
$T^{\{bb\}}_{\{\bar c\bar u\}}\to \overline D^0  B^- l^-\bar\nu $ & $ \left(a_2+a_3\right) V_{\text{ub}}$&
$T^{\{bb\}}_{\{\bar c\bar u\}}\to D^-  \overline B^0 l^-\bar\nu $ & $ \left(a_2+a_3\right) V_{\text{ub}}$\\\hline
$T^{\{bb\}}_{\{\bar c\bar u\}}\to  D^-_s  \overline B^0_s l^-\bar\nu $ & $ \left(a_2+a_3\right) V_{\text{ub}}$&&\\\hline
\hline
$T^{\{bb\}}_{\{\bar c\bar u\}}\to B^- l^-\bar\nu $ & $ a_5$&
$T^{\{bb\}}_{\{\bar c\bar d\}}\to \overline B^0 l^-\bar\nu $ & $ a_5$\\\hline
$T^{\{bb\}}_{\{\bar c\bar s\}}\to \overline B^0_s l^-\bar\nu $ & $ a_5$&&\\\hline
\hline
$T^{\{bb\}}_{\{\bar c\bar u\}}\to B^-  \pi^0  l^-\bar\nu $ & $ \frac{a_6}{\sqrt{2}}$&
$T^{\{bb\}}_{\{\bar c\bar u\}}\to B^-  \eta  l^-\bar\nu $ & $ \frac{a_6}{\sqrt{6}}$\\\hline
$T^{\{bb\}}_{\{\bar c\bar u\}}\to \overline B^0  \pi^-  l^-\bar\nu $ & $ a_6$&
$T^{\{bb\}}_{\{\bar c\bar u\}}\to \overline B^0_s  K^-  l^-\bar\nu $ & $ a_6$\\\hline
$T^{\{bb\}}_{\{\bar c\bar d\}}\to B^-  \pi^+  l^-\bar\nu $ & $ a_6$&
$T^{\{bb\}}_{\{\bar c\bar d\}}\to \overline B^0  \pi^0  l^-\bar\nu $ & $ -\frac{a_6}{\sqrt{2}}$\\\hline
$T^{\{bb\}}_{\{\bar c\bar d\}}\to \overline B^0  \eta  l^-\bar\nu $ & $ \frac{a_6}{\sqrt{6}}$&
$T^{\{bb\}}_{\{\bar c\bar d\}}\to \overline B^0_s  \overline K^0  l^-\bar\nu $ & $ a_6$\\\hline
$T^{\{bb\}}_{\{\bar c\bar s\}}\to B^-  K^+  l^-\bar\nu $ & $ a_6$&
$T^{\{bb\}}_{\{\bar c\bar s\}}\to \overline B^0  K^0  l^-\bar\nu $ & $ a_6$\\\hline
$T^{\{bb\}}_{\{\bar c\bar s\}}\to \overline B^0_s  \eta  l^-\bar\nu $ & $ -\sqrt{\frac{2}{3}} a_6$&&\\\hline
\hline
$T^{\{bb\}}_{\{\bar c\bar u\}}\to B^-  J/\psi l^-\bar\nu $ & $ a_7$&
$T^{\{bb\}}_{\{\bar c\bar d\}}\to \overline B^0  J/\psi l^-\bar\nu $ & $ a_7$\\\hline
$T^{\{bb\}}_{\{\bar c\bar s\}}\to \overline B^0_s  J/\psi l^-\bar\nu $ & $ a_7$&&\\\hline
\hline
$T^{\{bb\}}_{\{\bar c\bar u\}}\to  D^0  B_c^- l^-\bar\nu $ & $ a_8$&
$T^{\{bb\}}_{\{\bar c\bar d\}}\to  D^+  B_c^- l^-\bar\nu $ & $ a_8$\\\hline
$T^{\{bb\}}_{\{\bar c\bar s\}}\to  D^+_s  B_c^- l^-\bar\nu $ & $ a_8$&&\\\hline
\end{tabular}
\end{table}

For the semi-leptonic transition of $\bar c\to \bar d/\bar s \ell^- \overline \nu_{\ell}$, the Hamiltonian of mesonic decays is easy to construct, given as follows,
\begin{eqnarray}
&\mathcal{H}_{eff}&=c_1 (T^{\{bb\}}_{\{\bar c\bar q\}})_i (H_3)_j \overline{B}^i \overline{B}^j \bar \ell \nu.
\end{eqnarray}
The corresponding Feynman diagram is drawn in Fig.~\ref{fig:topology1}.(e). More technically, we can expand the Hamiltonian above and collect the decay amplitudes,
\begin{eqnarray*}
\mathcal{A}(T^{\{bb\}}_{\{\bar c\bar u\}}\to \overline B^0  B^- l^-\bar\nu )=c_1 V_{\text{cd}},\
\mathcal{A}(T^{\{bb\}}_{\{\bar c\bar u\}}\to \overline B^0_s  B^- l^-\bar\nu )=c_1 V_{\text{cs}},\
\mathcal{A}(T^{\{bb\}}_{\{\bar c\bar d\}}\to \overline B^0  \overline B^0 l^-\bar\nu )= 2 c_1 V_{\text{cd}},\\
\mathcal{A}(T^{\{bb\}}_{\{\bar c\bar d\}}\to \overline B^0  \overline B^0_s l^-\bar\nu )= c_1 V_{\text{cs}},\
\mathcal{A}(T^{\{bb\}}_{\{\bar c\bar s\}}\to \overline B^0_s  \overline B^0 l^-\bar\nu )= c_1 V_{\text{cd}},\
\mathcal{A}(T^{\{bb\}}_{\{\bar c\bar s\}}\to \overline B^0_s  \overline B^0_s l^-\bar\nu )= 2 c_1 V_{\text{cs}}.
\end{eqnarray*}
The relations  of  decay widths given as follows.
\begin{eqnarray*}
\Gamma(T^{\{bb\}}_{\{\bar c\bar u\}}\to B^-  \overline B^0 l^-\bar\nu)=\frac{1}{2}\Gamma(T^{\{bb\}}_{\{\bar c\bar d\}}\to \overline B^0  \overline B^0 l^-\bar\nu)=\Gamma(T^{\{bb\}}_{\{\bar c\bar s\}}\to \overline B^0  \overline B^0_s l^-\bar\nu),\\
\Gamma(T^{\{bb\}}_{\{\bar c\bar u\}}\to \overline B^0_s  B^- l^-\bar\nu)=
\Gamma(T^{\{bb\}}_{\{\bar c\bar d\}}\to \overline B^0  \overline B^0_s l^-\bar\nu)=
\frac{1}{2}\Gamma(T^{\{bb\}}_{\{\bar c\bar s\}}\to \overline B^0_s  \overline B^0_s l^-\bar\nu).
\end{eqnarray*}

\subsection{Non-leptonic two-body mesonic decays }
\label{sec:nonleptonic_decay}
\begin{figure}
  \centering
  \includegraphics[width=0.99\columnwidth]{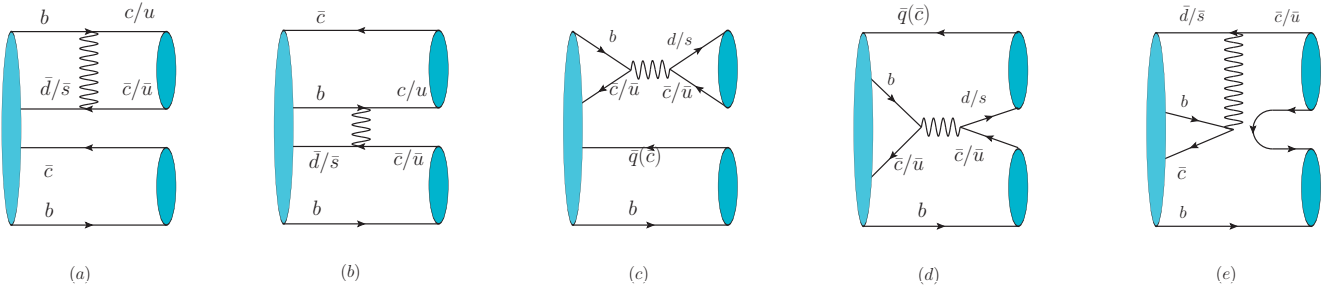}\\
  \caption{The Feynman diagrams for the tetraquark $T^{\{bb\}}_{\{\bar c\bar q\}}$ non-leptonic decays into mesons. }\label{fig:topology2}
\end{figure}

In the SU(3) flavor symmetry, the two-body mesonic decays Hamiltonian of the transition $b\to c\bar c d/s$  is constructed as
\begin{eqnarray}
&\mathcal{H}_{eff}&=a_1 (T^{\{bb\}}_{\{\bar c\bar q\}})_i (H_3)^i \overline{B}^j D_j+a_2 (T^{\{bb\}}_{\{\bar c\bar q\}})_i (H_3)^j \overline{B}^i D_j  +a_3 (T^{\{bb\}}_{\{\bar c\bar q\}})_i  (H_3)^j M^i_j \overline{B}_c\nonumber\\
&&+a_4 (T^{\{bb\}}_{\{\bar c\bar q\}})_i  (H_3)^i \overline{B}_c J/\psi.
\end{eqnarray}
We draw the corresponding Feynman diagrams shown in Fig.~\ref{fig:topology2}.(a-e). Particularly, the terms $a_1$ and $a_2$ with the final states $B$ plus $D$ are related to the diagrams Fig.~\ref{fig:topology2}.(c,e). The terms $a_4$ is corresponding with the diagrams Fig.~\ref{fig:topology2}.(a,b). In addition, the $a_3$ term with the light meson and $B_c$ meson in the final states is relevant with the Fig.~\ref{fig:topology2}.(d). Based on the Hamiltonian above, we get the decay amplitudes collected in Tab.~\ref{tab:bq_2mesons}. Consistently, the relations between different decay widths given as follows.
\begin{eqnarray*}
    \Gamma(T^{\{bb\}}_{\{\bar c\bar d\}}\to B^-\overline D^0)= { }\Gamma(T^{\{bb\}}_{\{\bar c\bar d\}}\to \overline B^0_s D^-_s), \Gamma(T^{\{bb\}}_{\{\bar c\bar s\}}\to B^-\overline D^0)= { }\Gamma(T^{\{bb\}}_{\{\bar c\bar s\}}\to \overline B^0D^-),\\
    \Gamma(T^{\{bb\}}_{\{\bar c\bar u\}}\to B^-D^-)= { }\Gamma(T^{\{bb\}}_{\{\bar c\bar s\}}\to \overline B^0_sD^-), \Gamma(T^{\{bb\}}_{\{\bar c\bar u\}}\to B^- D^-_s)= { }\Gamma(T^{\{bb\}}_{\{\bar c\bar d\}}\to \overline B^0 D^-_s),\\
    \Gamma(T^{\{bb\}}_{\{\bar c\bar u\}}\to B_c^-\pi^- )= 6\Gamma(T^{\{bb\}}_{\{\bar c\bar d\}}\to B_c^-\eta )=
\Gamma(T^{\{bb\}}_{\{\bar c\bar s\}}\to B_c^-K^0 )=2\Gamma(T^{\{bb\}}_{\{\bar c\bar d\}}\to B_c^-\pi^0 ),\\
      \Gamma(T^{\{bb\}}_{\{\bar c\bar u\}}\to B_c^-K^- )= \frac{3}{2}\Gamma(T^{\{bb\}}_{\{\bar c\bar s\}}\to B_c^-\eta )=
\Gamma(T^{\{bb\}}_{\{\bar c\bar d\}}\to B_c^-\overline K^0 ).
\end{eqnarray*}
\begin{table}
\caption{Tetraquark $T^{\{bb\}}_{\{\bar{c} \bar{q}\}}$ decays into two mesons with the transition of  $b\to c\bar c d/s$.}\label{tab:bq_2mesons}\begin{tabular}{|c|c|c|c|c|c|c|c}\hline\hline
channel & amplitude &channel &amplitude\\\hline
$T^{\{bb\}}_{\{\bar c\bar u\}}\to   B^-  D^- $ & $ a_2 V_{\text{cd}}^*$&
$T^{\{bb\}}_{\{\bar c\bar u\}}\to   B^-   D^-_s $ & $ a_2 V_{\text{cs}}^*$\\\hline
$T^{\{bb\}}_{\{\bar c\bar d\}}\to   B^-  \overline D^0 $ & $ a_1 V_{\text{cd}}^*$&
$T^{\{bb\}}_{\{\bar c\bar d\}}\to   \overline B^0  D^- $ & $ \left(a_1+a_2\right) V_{\text{cd}}^*$\\\hline
$T^{\{bb\}}_{\{\bar c\bar d\}}\to   \overline B^0   D^-_s $ & $ a_2 V_{\text{cs}}^*$&
$T^{\{bb\}}_{\{\bar c\bar d\}}\to   \overline B^0_s   D^-_s $ & $ a_1 V_{\text{cd}}^*$\\\hline
$T^{\{bb\}}_{\{\bar c\bar s\}}\to   B^-  \overline D^0 $ & $ a_1 V_{\text{cs}}^*$&
$T^{\{bb\}}_{\{\bar c\bar s\}}\to   \overline B^0  D^- $ & $ a_1 V_{\text{cs}}^*$\\\hline
$T^{\{bb\}}_{\{\bar c\bar s\}}\to   \overline B^0_s  D^- $ & $ a_2 V_{\text{cd}}^*$&
$T^{\{bb\}}_{\{\bar c\bar s\}}\to   \overline B^0_s   D^-_s $ & $ \left(a_1+a_2\right) V_{\text{cs}}^*$\\\hline
\hline
$T^{\{bb\}}_{\{\bar c\bar u\}}\to   B_c^-  \pi^-  $ & $ a_3 V_{\text{cd}}^*$&
$T^{\{bb\}}_{\{\bar c\bar u\}}\to   B_c^-  K^-  $ & $ a_3 V_{\text{cs}}^*$\\\hline
$T^{\{bb\}}_{\{\bar c\bar d\}}\to   B_c^-  \pi^0  $ & $ -\frac{a_3 V_{\text{cd}}^*}{\sqrt{2}}$&
$T^{\{bb\}}_{\{\bar c\bar d\}}\to   B_c^-  \overline K^0  $ & $ a_3 V_{\text{cs}}^*$\\\hline
$T^{\{bb\}}_{\{\bar c\bar d\}}\to   B_c^-  \eta  $ & $ \frac{a_3 V_{\text{cd}}^*}{\sqrt{6}}$&
$T^{\{bb\}}_{\{\bar c\bar s\}}\to   B_c^-  K^0  $ & $ a_3 V_{\text{cd}}^*$\\\hline
$T^{\{bb\}}_{\{\bar c\bar s\}}\to   B_c^-  \eta  $ & $ -\sqrt{\frac{2}{3}} a_3 V_{\text{cs}}^*$& &\\\hline
\hline
$T^{\{bb\}}_{\{\bar c\bar d\}}\to   B_c^-  J/\psi $ & $ a_4 V_{\text{cd}}^*$&
$T^{\{bb\}}_{\{\bar c\bar s\}}\to   B_c^-  J/\psi $ & $ a_4 V_{\text{cs}}^*$\\\hline
\hline
\end{tabular}
\end{table}

The hadronic level Hamiltonian of the transition of $b\to c\bar u d/s$ can be constructed as
\begin{eqnarray}
&\mathcal{H}_{eff}&=b_1' (T^{\{bb\}}_{\{\bar c\bar q\}})_i (H_8)^i_j \overline{B}^j J/\psi +b_2' (T^{\{bb\}}_{\{\bar c\bar q\}})_i (H_8)^i_j \overline{D}^j \overline{B}_c  +b_3' (T^{\{bb\}}_{\{\bar c\bar q\}})_i  (H_8)^i_j M^j_k \overline{B}^k \nonumber\\
&&+b_4' (T^{\{bb\}}_{\{\bar c\bar q\}})_i  (H_8)^k_j M^i_k \overline{B}^j +b_5' (T^{\{bb\}}_{\{\bar c\bar q\}})_i  (H_8)^k_j M^j_k \overline{B}^i.
\end{eqnarray}
The corresponding Feynman diagrams are shown in Fig.~\ref{fig:topology2}.(a-e). Further more, we expand the Hamiltonian and obtain the decay amplitudes collected in Tab.~\ref{tab:bq_B_Jpsi}. Besides, the relation between different decay channels can be deduced directly, given as follows.
\begin{eqnarray*}
    \Gamma(T^{\{bb\}}_{\{\bar c\bar s\}}\to \pi^0 B^-)= \frac{1}{2}\Gamma(T^{\{bb\}}_{\{\bar c\bar s\}}\to \pi^- \overline B^0).
\end{eqnarray*}

\begin{table}
\caption{Tetraquark $T^{\{bb\}}_{\{\bar{c} \bar{q}\}}$ decays into two mesons with the transition of  $b\to c\bar u d/s$ or $b\to u\bar c d/s$.}\label{tab:bq_B_Jpsi}\begin{tabular}{|c|c|c|c|c|c|c|c}\hline\hline
channel & amplitude &channel &amplitude\\\hline
$T^{\{bb\}}_{\{\bar c\bar d\}}\to   B^-  J/\psi $ & $ b_1 V_{\text{ud}}^*$&
$T^{\{bb\}}_{\{\bar c\bar s\}}\to   B^-  J/\psi $ & $ b_1 V_{\text{us}}^*$\\\hline
\hline
$T^{\{bb\}}_{\{\bar c\bar d\}}\to    D^0  B_c^- $ & $ b_2 V_{\text{ud}}^*$&
$T^{\{bb\}}_{\{\bar c\bar s\}}\to    D^0  B_c^- $ & $ b_2 V_{\text{us}}^*$\\\hline
\hline
$T^{\{bb\}}_{\{\bar c\bar u\}}\to   \pi^-   B^- $ & $ \left(b_4+b_5\right) V_{\text{ud}}^*$&
$T^{\{bb\}}_{\{\bar c\bar u\}}\to   K^-   B^- $ & $ \left(b_4+b_5\right) V_{\text{us}}^*$\\\hline
$T^{\{bb\}}_{\{\bar c\bar d\}}\to   \pi^0   B^- $ & $ \frac{\left(b_3-b_4\right) V_{\text{ud}}^*}{\sqrt{2}}$&
$T^{\{bb\}}_{\{\bar c\bar d\}}\to   \pi^-   \overline B^0 $ & $ \left(b_3+b_5\right) V_{\text{ud}}^*$\\\hline
$T^{\{bb\}}_{\{\bar c\bar d\}}\to   \overline K^0   B^- $ & $ b_4 V_{\text{us}}^*$&
$T^{\{bb\}}_{\{\bar c\bar d\}}\to   K^-   \overline B^0 $ & $ b_5 V_{\text{us}}^*$\\\hline
$T^{\{bb\}}_{\{\bar c\bar d\}}\to   K^-   \overline B^0_s $ & $ b_3 V_{\text{ud}}^*$&
$T^{\{bb\}}_{\{\bar c\bar d\}}\to   \eta   B^- $ & $ \frac{\left(b_3+b_4\right) V_{\text{ud}}^*}{\sqrt{6}}$\\\hline
$T^{\{bb\}}_{\{\bar c\bar s\}}\to   \pi^0   B^- $ & $ \frac{b_3 V_{\text{us}}^*}{\sqrt{2}}$&
$T^{\{bb\}}_{\{\bar c\bar s\}}\to   \pi^-   \overline B^0 $ & $ b_3 V_{\text{us}}^*$\\\hline
$T^{\{bb\}}_{\{\bar c\bar s\}}\to   \pi^-   \overline B^0_s $ & $ b_5 V_{\text{ud}}^*$&
$T^{\{bb\}}_{\{\bar c\bar s\}}\to   K^0   B^- $ & $ b_4 V_{\text{ud}}^*$\\\hline
$T^{\{bb\}}_{\{\bar c\bar s\}}\to   K^-   \overline B^0_s $ & $ \left(b_3+b_5\right) V_{\text{us}}^*$&
$T^{\{bb\}}_{\{\bar c\bar s\}}\to   \eta   B^- $ & $ \frac{\left(b_3-2 b_4\right) V_{\text{us}}^*}{\sqrt{6}}$\\\hline
\hline
$T^{\{bb\}}_{\{\bar c\bar u\}}\to   D^-  B_c^- $ & $ c_1 V_{\text{cd}}^*$&
$T^{\{bb\}}_{\{\bar c\bar u\}}\to    D^-_s  B_c^- $ & $ c_1 V_{\text{cs}}^*$\\\hline
$T^{\{bb\}}_{\{\bar c\bar d\}}\to   \overline D^0  B_c^- $ & $ -c_1 V_{\text{cd}}^*$&
$T^{\{bb\}}_{\{\bar c\bar s\}}\to   \overline D^0  B_c^- $ & $ -c_1 V_{\text{cs}}^*$\\\hline
\hline
\end{tabular}
\end{table}

We can easily construct the hadronic level Hamiltonian for the transition of $b\to u\bar c d/s$ in the SU(3) symmetry,
\begin{eqnarray}
&\mathcal{H}_{eff}&=c_1' (T^{\{bb\}}_{\{\bar c\bar q\}})_i (H_{\bar3}'')^{[ij]} D_j \overline{B}_c.
\end{eqnarray}
The corresponding Feynman diagrams are shown in Fig.~\ref{fig:topology2}.(a-d). In addition, the amplitudes obtained from the Hamiltonian are collected in Tab.~\ref{tab:bq_B_Jpsi}, and the relations between different decay channels can be deduced as
\begin{eqnarray*}
    \Gamma(T^{\{bb\}}_{\{\bar c\bar d\}}\to \overline D^0B_c^-)= { }\Gamma(T^{\{bb\}}_{\{\bar c\bar u\}}\to D^-B_c^-),
     \Gamma(T^{\{bb\}}_{\{\bar c\bar s\}}\to \overline D^0B_c^-)= { }\Gamma(T^{\{bb\}}_{\{\bar c\bar u\}}\to  D^-_sB_c^-).
\end{eqnarray*}


Following the SU(3) analysis, the hadronic level Hamiltonian for the charmless transition of $b\to q_1\bar q_2 q_3$ can be constructed as
\begin{eqnarray}
&\mathcal{H}_{eff}&=d_1' (T^{\{bb\}}_{\{\bar c\bar q\}})_i (H_3)^{j} M^i_j \overline{B}_c + d_2' (T^{\{bb\}}_{\{\bar c\bar q\}})_i (H_{\bar6})^{[ik]}_j M^j_k \overline{B}_c + d_3' (T^{\{bb\}}_{\{\bar c\bar q\}})_i (H_{15})^{\{ik\}}_j M^j_k \overline{B}_c\nonumber\\
&&+d_4' (T^{\{bb\}}_{\{\bar c\bar q\}})_i (H_3)^{i} D_j \overline{B}^j +d_5' (T^{\{bb\}}_{\{\bar c\bar q\}})_i (H_3)^{j} D_j \overline{B}^i +d_6' (T^{\{bb\}}_{\{\bar c\bar q\}})_i (H_6)^{[ik]}_j D_k \overline{B}^j \nonumber\\
&&+d_7' (T^{\{bb\}}_{\{\bar c\bar q\}})_i (H_{15})^{\{ik\}}_j D_k \overline{B}^j.
\end{eqnarray}
We expand the Hamiltonian and collect the decay amplitudes given in Tab.~\ref{tab:bq_Bc_Md}. In particular, there is no relation between different decay channels.
\begin{table}
\caption{Tetraquark $T^{\{bb\}}_{\{\bar{c} \bar{q}\}}$ decays into $B_c$ plus light meson or anti-charm plus B meson induced by the charmless transition $b\to d$ or $b\to s$.}\label{tab:bq_Bc_Md}\begin{tabular}{|c|c|c|c|c|c|c|c}\hline\hline
channel & amplitude &channel &amplitude\\\hline
$T^{\{bb\}}_{\{\bar c\bar u\}}\to   B_c^-  \pi^-  $ & $ d_1+d_2+3 d_3$&
$T^{\{bb\}}_{\{\bar c\bar d\}}\to   B_c^-  \pi^0  $ & $ -\frac{d_1+d_2-5 d_3}{\sqrt{2}}$\\\hline
$T^{\{bb\}}_{\{\bar c\bar d\}}\to   B_c^-  \eta  $ & $ \frac{d_1-3 d_2+3 d_3}{\sqrt{6}}$&
$T^{\{bb\}}_{\{\bar c\bar s\}}\to   B_c^-  K^0  $ & $ d_1-d_2-d_3$\\\hline
$T^{\{bb\}}_{\{\bar c\bar u\}}\to   B^-  D^- $ & $ d_5+d_6+3 d_7$&
$T^{\{bb\}}_{\{\bar c\bar d\}}\to   B^-  \overline D^0 $ & $ d_4-d_6+3 d_7$\\\hline
$T^{\{bb\}}_{\{\bar c\bar d\}}\to   \overline B^0  D^- $ & $ d_4+d_5-2 d_7$&
$T^{\{bb\}}_{\{\bar c\bar d\}}\to   \overline B^0_s   D^-_s $ & $ d_4+d_6-d_7$\\\hline
$T^{\{bb\}}_{\{\bar c\bar s\}}\to   \overline B^0_s  D^- $ & $ d_5-d_6-d_7$&&\\\hline
\hline

$T^{\{bb\}}_{\{\bar c\bar u\}}\to   B_c^-  K^-  $ & $ d_1+d_2+3 d_3$&
$T^{\{bb\}}_{\{\bar c\bar d\}}\to   B_c^-  \overline K^0  $ & $ d_1-d_2-d_3$\\\hline
$T^{\{bb\}}_{\{\bar c\bar s\}}\to   B_c^-  \pi^0  $ & $ -\sqrt{2} \left(d_2-2 d_3\right)$&
$T^{\{bb\}}_{\{\bar c\bar s\}}\to   B_c^-  \eta  $ & $ -\sqrt{\frac{2}{3}} \left(d_1-3 d_3\right)$\\\hline
$T^{\{bb\}}_{\{\bar c\bar u\}}\to   B^-   D^-_s $ & $ d_5+d_6+3 d_7$&
$T^{\{bb\}}_{\{\bar c\bar d\}}\to   \overline B^0   D^-_s $ & $ d_5-d_6-d_7$\\\hline
$T^{\{bb\}}_{\{\bar c\bar s\}}\to   B^-  \overline D^0 $ & $ d_4-d_6+3 d_7$&
$T^{\{bb\}}_{\{\bar c\bar s\}}\to   \overline B^0  D^- $ & $ d_4+d_6-d_7$\\\hline
$T^{\{bb\}}_{\{\bar c\bar s\}}\to   \overline B^0_s   D^-_s $ & $ d_4+d_5-2 d_7$&&\\\hline
\hline
\end{tabular}
\end{table}

For the charm quark decay $\bar c\to \bar q_1 q_2 \bar q_3$, the Hamiltonian of hadronic level can be easily constructed as
\begin{eqnarray}
&\mathcal{H}_{eff}&=f_1 (T^{\{bb\}}_{\{\bar c\bar q\}})_i (H_{\overline{15}})_{\{jk\}}^{i} \overline B^j \overline B^k.
\end{eqnarray}
Expanding the Hamiltonian above, one can obtain the decay amplitudes given as follows,
\begin{eqnarray*}
&&\mathcal{A}(T^{\{bb\}}_{\{\bar c\bar d\}}\to   B^-  \overline B^0_s)=   2 f_1,
\mathcal{A}(T^{\{bb\}}_{\{\bar c\bar d\}}\to   B^-  \overline B^0)=-2 f_1 \sin\theta_C,
\mathcal{A}(T^{\{bb\}}_{\{\bar c\bar s\}}\to   B^-  \overline B^0_s)=2 f_1 \sin\theta_C,\\
&&\mathcal{A}(T^{\{bb\}}_{\{\bar c\bar s\}}\to   B^-  \overline B^0)= 2 f_1 {\sin\theta_C}^2.
\end{eqnarray*}
Consistently, the relation between different decay channels is deduced as
\begin{eqnarray*}
\Gamma(T^{\{bb\}}_{\{\bar c\bar d\}}\to   B^-  \overline B^0)=
\Gamma(T^{\{bb\}}_{\{\bar c\bar s\}}\to   B^-  \overline B^0_s).
\end{eqnarray*}

\section{Golden Decay Channels}
\label{sec:golden_channels}

In order to reconstruct the tetraquark $T_{\{\bar c \bar q\}}^{\{bb\}}$ experimentally, we will discuss the golden channels in the section.
In the discussion given in the previous sections,  the  final meson can  be replaced by its corresponding counterpart, which has the same quark constituent but different quantum numbers. For instance, we can replace  $\pi^0$ by $\rho^0$, $K^+$ by $K^{*+}$.

Generally, the weak decays are closely related with the CKM element. In addition, the charged particles are easy to detect than neutral one at hadron colliders like LHC. Therefore, one choose the following criteria for the Golden decay  channels.
\begin{itemize}
\item Branching fractions:  One should choose the cabibbo allowed decay modes $c\to s \ell^+ \nu_{\ell}$ for the charm quark semi-leptonic decays and $c\to s \bar d u$ for the non-leptonic decays. For bottom quark decays, the semi-leptonic decays $b\to c \ell^- \overline \nu_{\ell}$ and non-leptonic decays $b\to c\bar ud$ or $b\to c\bar cs$ give the largest branching fractions.

\item Detection efficiency:  Since it is easy to detect the charged particles, one remove the channels with   $\pi^0$, $\eta$, $\phi$, $\rho^{\pm}(\to \pi^{\pm}\pi^0$), $K^{*\pm}(\to K^{\pm}\pi^0$) and $\omega$, but keep the modes with $\pi^\pm, K^0(\to \pi^+\pi^-), \rho^0(\to \pi^+\pi^-)$.

\end{itemize}


\begin{table}
 \caption{Cabibbo allowed ${b\bar c}{b\bar q}$  decays.  }\label{tab:T_golden_meson_c}\begin{tabular}{|c  c   c   c c|}\hline\hline
&&$b$ quark decays&&\\\hline
$T^{\{bb\}}_{\{\bar c\bar u\}}\to B^- l^-\bar\nu $&$T^{\{bb\}}_{\{\bar c\bar d\}}\to \overline B^0 l^-\bar\nu $&$T^{\{bb\}}_{\{\bar c\bar s\}}\to \overline B^0_s l^-\bar\nu $&$T^{\{bb\}}_{\{\bar c\bar u\}}\to B^-  J/\psi l^-\bar\nu $&$T^{\{bb\}}_{\{\bar c\bar d\}}\to \overline B^0  J/\psi l^-\bar\nu $\\
$T^{\{bb\}}_{\{\bar c\bar s\}}\to \overline B^0_s  J/\psi l^-\bar\nu $&$T^{\{bb\}}_{\{\bar c\bar u\}}\to \overline B^0  \pi^-  l^-\bar\nu $&$T^{\{bb\}}_{\{\bar c\bar u\}}\to \overline B^0_s  K^-  l^-\bar\nu $&$T^{\{bb\}}_{\{\bar c\bar d\}}\to B^-  \pi^+  l^-\bar\nu $&$T^{\{bb\}}_{\{\bar c\bar d\}}\to \overline B^0_s  \overline K^0  l^-\bar\nu $\\
$T^{\{bb\}}_{\{\bar c\bar s\}}\to B^-  K^+  l^-\bar\nu $&$T^{\{bb\}}_{\{\bar c\bar s\}}\to \overline B^0  K^0  l^-\bar\nu $&$T^{\{bb\}}_{\{\bar c\bar u\}}\to  D^0  B_c^- l^-\bar\nu$&$T^{\{bb\}}_{\{\bar c\bar d\}}\to  D^+  B_c^- l^-\bar\nu$&$T^{\{bb\}}_{\{\bar c\bar s\}}\to  D^+_s  B_c^- l^-\bar\nu$,
\\
$T^{\{bb\}}_{\{\bar c\bar u\}}\to   B^-   D^-_s $&$T^{\{bb\}}_{\{\bar c\bar d\}}\to   \overline B^0   D^-_s $&
$T^{\{bb\}}_{\{\bar c\bar s\}}\to   B^-  \overline D^0 $ &$T^{\{bb\}}_{\{\bar c\bar s\}}\to   \overline B^0  D^- $ &$T^{\{bb\}}_{\{\bar c\bar s\}}\to   \overline B^0_s   D^-_s $\\
$T^{\{bb\}}_{\{\bar c\bar u\}}\to   B_c^-  K^-  $&
$T^{\{bb\}}_{\{\bar c\bar d\}}\to   B_c^-  \overline K^0  $&$T^{\{bb\}}_{\{\bar c\bar s\}}\to   B_c^-  J/\psi $ &$T^{\{bb\}}_{\{\bar c\bar d\}}\to   B^-  J/\psi $ &$T^{\{bb\}}_{\{\bar c\bar d\}}\to    D^0  B_c^- $\\
$T^{\{bb\}}_{\{\bar c\bar u\}}\to   \pi^-   B^- $&$T^{\{bb\}}_{\{\bar c\bar d\}}\to   \pi^-   \overline B^0 $&$T^{\{bb\}}_{\{\bar c\bar d\}}\to   K^-   \overline B^0_s $ &$T^{\{bb\}}_{\{\bar c\bar s\}}\to   \pi^-   \overline B^0_s $&$T^{\{bb\}}_{\{\bar c\bar s\}}\to   K^0   B^- $\\
\hline
&&$\bar c$ quark decays&&\\\hline
$T^{\{bb\}}_{\{\bar c\bar u\}}\to \overline B^0_s  B^- l^-\bar\nu $&$T^{\{bb\}}_{\{\bar c\bar d\}}\to \overline B^0_s  \overline B^0 l^-\bar\nu $&$T^{\{bb\}}_{\{\bar c\bar s\}}\to \overline B^0_s  \overline B^0_s l^-\bar\nu $&
$T^{\{bb\}}_{\{\bar c\bar d\}}\to   B^-  \overline B^0_s $&\\
\hline\hline
\end{tabular}
\end{table}

We collect the golden channels with the two-body non-leptonic decays and  three- or four-body semi-leptonic decays for the triply heavy tetraquark  $T_{\{\bar c \bar q\}}^{\{bb\}}$, given in Tab.~\ref{tab:T_golden_meson_c}.

Some comments are appropriate in the following. For the charm quark decays, the typical branching fraction is at a few percent level. In addition, the reconstruction of final states, such as bottom mesons, is at the order $10^{-3}$ or even smaller on the experiment side. Therefore the branching fraction for the charm quark decays chains to reconstruct the $T_{\{\bar c \bar q\}}^{\{bb\}}$ might be at the order $10^{-5}$, or smaller. For the bottom quark decays, the typical branching fraction is at the order $10^{-3}$. The reconstruction of final states  $J/\psi$ or $D$ meson would introduce  another factor  $10^{-3}$. Therefore the branching fraction of these channels could be at  the order $10^{-5}$.


\section{Conclusions}
\label{sec:conclusions}

In the paper, we have studied the lifetimes and weak decays of the heavy tetraquarks ${b\bar c}{b\bar q}$. With the help of heavy quark expanding(HQE), we calculated the lifetimes of ${b\bar c}{b\bar q}$ at the leading order and next-to-leading order. Particularly, the lifetime at NLO is $\tau(T^{\{bb\}}_{\{\bar c\bar q\}})=(0.70\pm0.09)\times 10^{-12} \ s$. In addition , we discuss the weak decays of ${b\bar c}{b\bar q}$ under the  SU(3) analysis which has been successfully applied into heavy meson and baryon processes. Following the building blocks in SU(3) analysis, we constructed the hadronic level Hamiltonian. More technically, the non-perturbative effects are collected into parameters, such as $a_{i},b_{j},...$. According to the Hamiltonian, one can derive the relations between different decay channels. Finally, we analyzed the semi-leptonic and non-leptonic mesonic decays of tetraquarks ${b\bar c}{b\bar q}$, and made a collection of the golden channels for the searching of triply heavy tetraquarks ${b\bar c}{b\bar q}$ in further experiments.

\section*{Acknowledgments}

We thank Prof. Wei Wang for invaluable advice and  helpful discussions throughout the project.

\end{document}